\documentclass[manuscript]{emulateapj}
\slugcomment{accepted by ApJ}

\shorttitle{The oldest X-ray supernovae}
\shortauthors{Soria \& Perna}

\begin{document}


\title{The oldest X-ray supernovae: X-ray emission 
from 1941C, 1959D, 1968D}


\author{Roberto Soria}
\affil{MSSL, University College London, Holmbury House, Dorking RH5 6NT, 
  United Kingdom}

%\author{C. D. Biemesderfer\altaffilmark{4,5}}
%\affil{National Optical Astronomy Observatories, Tucson, AZ 85719}
%\email{aastex-help@aas.org}

\and

\author{Rosalba Perna}
\affil{JILA and Department of Astrophysical and Planetary Sciences, 
University of Colorado, Boulder, CO 80309, USA}


%\altaffiltext{1}{Visiting Astronomer, Cerro Tololo Inter-American Observatory.
%CTIO is operated by AURA, Inc.\ under contract to the National Science
%\altaffiltext{2}{Society of Fellows, Harvard University.}
%\altaffiltext{3}{present address: Center for Astrophysics,
%    60 Garden Street, Cambridge, MA 02138}
%\altaffiltext{4}{Visiting Programmer, Space Telescope Science Institute}
%\altaffiltext{5}{Patron, Alonso's Bar and Grill}


\begin{abstract}
We have studied the X-ray emission from four historical 
Type-II supernovae (the newly-recovered 1941C in NGC\,4631  
and 1959D in NGC\,7331; and 1968D, 1980K in NGC\,6946), 
using {\it Chandra} ACIS-S imaging. In particular, the first three 
are the oldest ever found in the X-ray band, and provide constraints 
on the properties of the stellar wind and circumstellar matter encountered 
by the expanding shock at more advanced stages 
in the transition towards the remnant phase. 
We estimate emitted luminosities $\approx 5 \times 10^{37}$ 
erg s$^{-1}$ for SN\,1941C, $\sim$ a few $\times 10^{37}$ 
erg s$^{-1}$ for SN\,1959D, $\approx 2 \times 10^{38}$ 
erg s$^{-1}$ for SN\,1968D, and $\approx 4 \times 10^{37}$ 
erg s$^{-1}$ for SN\,1980K, in the $0.3$--$8$ keV band.
X-ray spectral fits to SN\,1968D suggest the presence of a harder component, 
possibly a power law with photon index $\approx 2$, 
contributing $\approx 10^{37}$ erg s$^{-1}$ in the $2$--$10$ keV band.
We speculate that it may be evidence of non-thermal emission 
from a Crab-like young pulsar.
\end{abstract}





\section{Introduction}

Circumstellar interaction models \citep{Fr96,IL03}
predict two characteristic phases in the evolution of a core-collapse supernova (SN) 
over timescales of a few decades. In the first few years, the outgoing shock front 
is expanding through the material deposited by stellar winds from the 
progenitor star. During this phase, the radio and X-ray 
emission decline at a rate $\sim t^{-s}$ with $s \sim 1$--$4$ 
\citep{IL03,IK05,Sto06}.
After a few decades, the shock reaches the boundary 
with the interstellar medium (SN remnant phase), 
and the luminosity decline flattens. Thus, studying the evolution 
of an ageing SN, as it evolves towards the remnant stage, can 
provide information on the final few thousand years of evolution 
of the progenitor star.
Characteristic temperatures of the shocked gas are $\sim 0.5$--$1$ keV, 
leading to emission in the soft X-ray band.

Soft thermal emission may not be the only source of X-rays 
in an ageing SN. A large fraction of core-collapse SNe 
(about 87\% for a solar
metallicity and a Salpeter initial mass function; Heger et al.~2003)
is expected to leave behind a spinning, isolated neutron star.
Observations of known pulsars show that a fraction of the rotational
energy loss $\dot{E}_{\rm rot}$ is converted into X-ray radiation
\citep{BT97,P02}
%(Becker \& Trumper 1997; Possenti et al. 2002),
making many neutron stars visible as young X-ray pulsars, after the central
region becomes optically thin (a few decades; Chevalier \& Fransson 1994). 
If most pulsars had millisecond periods at birth, they would constitute
a substantial fraction of the population of X-ray point sources 
in star-forming galaxies \citep{PS04}.  
Conversely, 
limits on the X-ray luminosities of a sizeble sample of young pulsars can be
used to constrain their initial $\dot{E}_{\rm rot}$, and hence
the distribution of their spin periods at birth \citep{PS08}.
Since very young SNe have a substantial optical
depth, for these studies it is especially important to gather a sample
of ageing SNe, and measure or constrain the X-ray luminosity 
of the pulsars possibly created in those events.

Only 35 core-collapse SNe have been found in X-rays so 
far\footnote{http://lheawww.gsfc.nasa.gov/users/immler/supernovae\_list.html}; 
and there is still a big gap 
between the oldest historical SNe and the youngest remnants. 
%Finding more SNe in this age range would
%both improve the statistics for pulsar studies, as well as allow
%to study the evolutionary phases between the SN and the SNR. 
In this {\em Letter} we report our study of X-ray emission from
four historical SNe, using {\em Chandra X-ray Observatory} data:
SN\,1941C, SN\,1959D, SN\,1968D and SN\,1980K. The first 3 
are the oldest ever found in the X-ray band.

\begin{figure*}
\begin{center}
\includegraphics[angle=0,scale=.75]{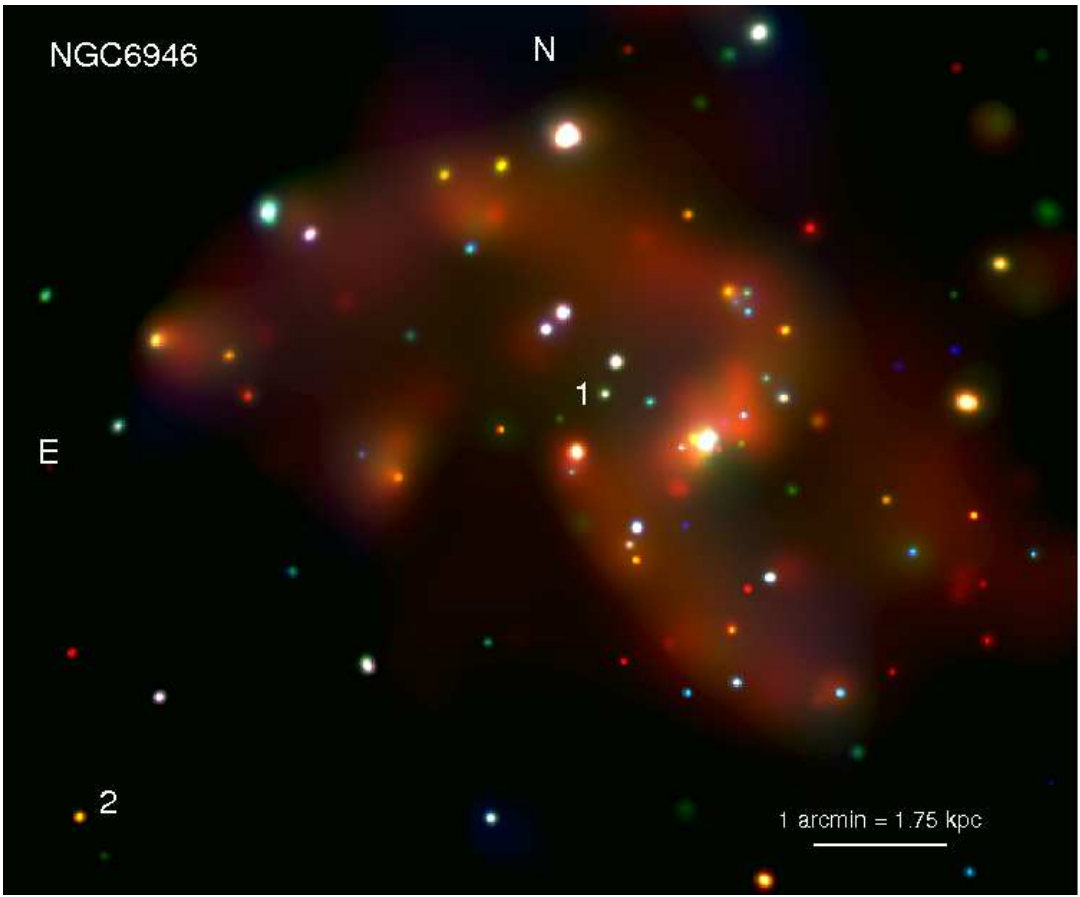}
\end{center}
\caption{Smoothed true-color image of NGC\,6946, 
from the combined {\it Chandra} ACIS-S observations. 
Red = $0.3$--$1$ keV; green = $1$--$2$ keV; blue = $2$--$8$ keV. 
SN\,1968D is labelled as '1';  SN\,1980K is labelled as '2' 
(note the softer color).}
\end{figure*}

\begin{deluxetable*}{lccccccr}
\tabletypesize{\scriptsize}
%\rotate
\tablecaption{Log of the observations, and fitted X-ray positions of the three oldest X-ray SNe \label{tbl-1}}
\tablewidth{0pt}
\tablehead
{
%\colhead{\,}&\colhead{\ }&\colhead{}&\colhead{}
%       &\colhead{\ }&\colhead{}&\colhead{}&\colhead{}\\[8pt]
%       & \multicolumn{5}{c}{Rate ($10^{-4}$ ACIS-S ct s$^{-1}$)}\\[2pt]
       \colhead{Galaxy}        & 
        \colhead{Distance}&
        \colhead{SN ID} &
        \colhead{RA}& 
        \colhead{Dec} & \colhead{Obs ID} & \colhead{Date} & \colhead{Live time}\\[2pt]
& (Mpc) &&&&&& (ks) }
\startdata
NGC\,4136 & 10 & 1941C  & 12:09:21.01 & $+$29:54:32.5  & 2920  & 2002-03-07 & 18.5 \\[2pt]
 &   &   &  &  & 2921  & 2002-06-08 & 19.7 \\[2pt]
NGC\,7331 & 15 & 1959D  & 22:37:01.82 & $+$34:25:08.3  & 2198  & 2001-01-27 & 29.5 \\[2pt]
NGC\,6946 & 6  & 1968D  & 20:34:58.40 & $+$60:09:34.4  & 1043 & 2001-09-07 & 58.3 \\[2pt]
 &   &   &  &   & 4404 &  2002-11-25  & 30.0 \\[2pt]
 &   &   &  &   & 4631 &  2004-10-22  & 29.7 \\[2pt]
 &   &   &  &   & 4632 &  2004-11-06  & 28.0 \\[2pt]
 &   &   &  &   & 4633 &  2004-12-03  & 26.6 \\[-5pt]
\enddata
\end{deluxetable*}

\section{Data analysis}

We searched the {\it Chandra} public archive, and 
downloaded the observations that included in the field of view 
the location of historical core-collapse SNe older than 1970G 
(the oldest X-ray SN listed in Immler's catalog, which  
had overlooked SN\,1968D; see Section 3.3). We limited our search 
to host galaxies at distances $\la 15$ Mpc. The optical position 
of the SNe was taken from the CfA List of 
Supernovae\footnote{http://cfa-www.harvard.edu/iau/lists/Supernovae.html}, 
the Asiago Catalogue\footnote{http://web.pd.astro.it/supern/snean.txt} 
and the Sternberg Catalogue \citep{tsv04}. 
We estimate that optical positions have uncertainties between 
$\approx$ 1\arcsec and 2\arcsec. We used the Chandra Interactive 
Analysis of Observations ({\small CIAO}) 
software package (Version 3.4) to extract and examine 
the images in the full $0.3$--$8$ keV band, as well as 
in narrower bands. In particular, we used the {\it wavdetect} 
source finding routine. We found point-like X-ray 
sources associated with the SN position in three cases: 
SN\,1941C in NGC\,4136; SN\,1959D in NGC\,7331; and SN\,1968D 
in NGC\,6946, where we also recovered SN\,1980K. 
We then studied those systems in more detail.

NGC\,4136 was observed by {\it Chandra}/ACIS-S twice: on 2002 March 07 (ObsID 2920), 
for a live time of 18.5 ks, and on 2002 June 08 (ObsID 2921), 
for a live time of 19.7 ks. The level-2 event files in the public archive 
were created in 2006 with the standard data processing version DS7.6.8 
(i.e., with recent calibration data), and did not require reprocessing 
or astrometric corrections.
We defined a circular source region with radius 2\farcs5 (comprising 
95\% of the source counts at 2 keV, on axis), centered 
on the X-ray source, within 1\arcsec\ from the approximative 
optical position of SN\,1941C. There are no other X-ray sources in a radius 
of 25\arcsec, so a chance coincidence is very unlikely.
We extracted the background from an annulus of radii 
4\arcsec\ and 10\arcsec. We used the {\small CIAO} task {\it psextract} 
to create the source file, and {\it mkacisrmf} for the response. 
We recalculated the auxiliary response file with {\it mkarf}.
Considering the small number of detected counts, and the proximity 
of the two observations compared with the age of the SN, 
we coadded the two spectra for subsequent flux estimates, 
using a local code \citep{PNS03}.

NGC\,7331 was observed by {\it Chandra}/ACIS-S on 2001 January 27 (ObsID 2198),
for a live time of 29.5 ks. The level-2 event file from the archive 
was created with the standard data processing version DS7.6.8 
and did not require reprocessing. There is a faint source 
at $\approx 3\sigma$ significance at the position 
of SN\,1959D ($\la$ 1\arcsec\ from the optical position) and no other 
sources within $\approx$ 10\arcsec. We centered our source and background 
extraction regions around the X-ray position, and applied the {\small CIAO} 
tasks mentioned above.

NGC\,6946 was observed by {\it Chandra}/ACIS-S on 5 occasions 
between 2001 September 07 and 2004 December 03 (see Table 1 for details). 
The combined live time is 172.6 ks. The level-2 event files in the archive 
were produced with recent versions of the calibration files 
(DS 7.6.7.1 and DS7.6.8) and did not need reprocessing. 
A moderately bright source is detected at high significance 
at the position of SN\,1968D in each of the observations;  
SN\,1980K is found in the last four observations (it is outside 
the field of view in the first one). We extracted the source spectra, 
with associated background, response and auxiliary response files,
as explained above. We then studied both the individual spectra  
and a coadded one from all observations.

Finally, after we extracted all the spectra, we used {\small XSPEC} 
Version 12 \citep{Ar96} to do spectral fitting (when possible) or estimate 
flux rates in different bands.

\begin{table}
\begin{center}
\caption{Best-fit parameters for the coadded X-ray spectrum 
of SN\,1968D. Spectral model:
{\tt wabs*wabs*(po+ray)}. Values in brackets were 
%{\tt wabs$_{\rm Gal}$*wabs*(po+ray)}. Values in brackets were kept
fixed. Errors are 90\%
confidence levels for 1 interesting parameter ($\Delta {\rm C} =
2.7$). We also list the best-fit $\chi^2$ for the 3 models.
\label{tbl-3}}
%\vspace{0.2cm}
\begin{tabular}{l@{\ \ \ }c@{\ \ \ }c@{\ \ \ }r}
%\vspace{0.3cm}
\tableline\tableline\\
Parameter & Model 1 & Model 2 & Model 3\\
 & Value & Value & Value\\
\tableline\\
$N_{\rm {H,Gal}}$\tablenotemark{a} & $(1.9 \times 10^{21})$ & $(1.9 \times 10^{21})$
                & $(1.9 \times 10^{21})$\\[6pt]
$N_{\rm {H}}$ & $4.8^{+2.5}_{-2.9} \times 10^{21}$ & $10.7^{+2.2}_{-3.1} \times 10^{21}$ 
                & $11.1^{+2.8}_{-8.3} \times 10^{21}$\\[6pt]
$\Gamma$\tablenotemark{b}  & $3.4^{+0.4}_{-0.9}$ & - & $1.9^{+0.9}_{-1.6}$\\[6pt]
$N_{\rm {pl}}$\tablenotemark{c} &  $4.4^{+1.5}_{-2.4} \times 10^{-6}$
        & - & $0.7^{+0.6}_{-0.5} \times 10^{-6}$ \\[6pt]
$kT_{\rm{gas}}$ (keV) & -  & $0.69^{+0.11}_{-0.11}$  & $0.45^{+0.18}_{-0.12}$ \\[6pt]
$Z(Z_{\odot})$ & - & $(1.0)$ & $(1.0)$\\[6pt]
%$Z_{\rm 0-Ca} (Z_{\odot})$ & $(1.0)$ & $(1.0)$\\[6pt]
%$Z_{\rm Fe,Ni} (Z_{\odot})$ & $(1.0)$ & $1.6^{+1.0}_{-0.6}$\\[6pt]
$K_{\rm{rs}}$\tablenotemark{d} & - 
         & $1.1^{+0.7}_{-0.4} \times 10^{-5}$  & $1.7^{+1.1}_{-1.1} \times 10^{-5}$\\[6pt]
\tableline\\
C-stat/dof   & $14.2/10$ & $13.0/10$ & $6.6/8$\\[6pt]
$\chi^2$/dof & $11.4/10$ & $10.7/10$  & $6.4/8$\\[3pt]
\tableline\\
$f_{\rm x, 0.3-8}$\tablenotemark{e} &$3.3^{+0.9}_{-0.6}$
        & $2.5^{+0.2}_{-0.2}$   & $3.6^{+0.8}_{-1.6}$\\[6pt]
$L_{\rm x, 0.3-8}$\tablenotemark{f} &$11.5^{+20.8}_{-7.3}$
        & $15.1^{+9.0}_{-8.1}$   & $24.4^{+53.1}_{-20.7}$\\[6pt]
$L_{\rm x, 0.3-2}$\tablenotemark{g} & $10.8^{+21.1}_{-7.5}$
        & $14.7^{+9.0}_{-8.1}$  & $23.5^{+53.2}_{-20.7}$ \\[6pt]
$L_{\rm x, 2-10}$\tablenotemark{h} &$0.7^{+0.4}_{-0.2}$
        & $0.4^{+0.1}_{-0.1}$  & $1.1^{+0.5}_{-0.5}$\\[6pt]
\tableline\\[-12pt]
\end{tabular}
\tablenotetext{a}{From \citet{K05}. Units of cm$^{-2}$.}
\tablenotetext{b}{Photon index.}
\tablenotetext{c}{Units of photons keV$^{-1}$ cm$^{-2}$ s$^{-1}$, at 1 keV.}
\tablenotetext{d}{Raymond-Smith model normalization $K_{\rm rs} =
10^{-14}/\{4\pi\,[d_A\,(1+z)]^2\}\,
\int n_e n_H {\rm{d}}V$, where $d_A$ is
          the angular size distance to the source (cm), $n_e$ is the electron
          density (cm$^{-3}$), and $n_H$ is the hydrogen density (cm$^{-3}$).}
\tablenotetext{e}{Observed flux in the $0.3$--$8$ keV band; 
          units of $10^{-15}$ erg cm$^{-2}$ s$^{-1}$.}
\tablenotetext{f}{Unabsorbed luminosity in the $0.3$--$8$ keV band; 
          units of $10^{37}$ erg s$^{-1}$.}
\tablenotetext{g}{Unabsorbed luminosity in the $0.3$--$2$ keV band; 
          units of $10^{37}$ erg s$^{-1}$.}
\tablenotetext{h}{Unabsorbed luminosity in the $2$--$10$ keV band; 
          units of $10^{37}$ erg s$^{-1}$.}\\[2pt]
\end{center}
\end{table}

\section{Results}

\subsection{SN\,1941C}

SN\,1941C is a Type-II SN (no further sub-classification 
available) discovered on 1941 April 16 \citep{J41} in the face-on SBc 
galaxy NGC\,4136. {\it Chandra} observed it $\approx 61$ yr later. 
From the combined 38-ks 
observation, we estimate a count rate of $(6.5 \pm 1.3) \times 10^{-4}$ 
ct s$^{-1}$ in the $0.3$--$8$ keV band.  
With only $\approx 25$ net counts, we cannot do much spectral 
fitting; however, we have enough counts to find that the emission 
is not entirely in the soft band. We estimate a count rate of 
$(1.6 \pm 0.7) \times 10^{-4}$ in the $1.5$--$8$ keV band 
(i.e., about 6 out of 25 background-subtracted counts).
Assuming a fixed line-of-sight absorption of $1.5 \times 10^{20}$ 
cm$^{-2}$ \citep{K05}, 
and using Cash statistics \citep{Cash79}, we find that the X-ray colors 
are roughly consistent with a thermal plasma at $\approx 3$ keV 
(with a 90\% confidence limit of $kT > 1.2$ keV), or with a power-law 
spectrum with photon index $\Gamma = 1.8^{+0.6}_{-0.5}$.

The distance to the host galaxy is $9.7$ Mpc \citep{T88}, or 
$(8 \pm 2)$ Mpc \citep{GZK03}.
For the thermal-plasma model, the unabsorbed luminosity 
of SN\,1941C in the $0.3$--$8$ keV band is 
$4.6^{+2.7}_{-2.4} \times 10^{37} d_{10}$ erg s$^{-1}$, 
where $d_{10} = (d/10 {\rm ~Mpc})$. For the power-law model, 
the unabsorbed luminosity is 
$5.3^{+2.8}_{-2.0} \times 10^{37} d_{10}$ erg s$^{-1}$ 
in the $0.3$--$8$ keV band, and 
$3.1^{+2.2}_{-1.5} \times 10^{37} d_{10}$ erg s$^{-1}$
in the $2$--$10$ keV band.

\subsection{SN\,1959D}

SN\,1959D is a Type-IIL SN discovered on 1959 June 28 \citep{Hu59}
in the Sbc galaxy NC\,7331 ($d=15.1$ Mpc: Hughes et al.~1998). 
{\it Chandra} observed it at an age of $\approx 42$ yr.
We found an excess of ACIS-S counts at the SN position: 
7 total cts in the $0.3$--$8$ keV band from 
the source extraction region, compared with 
an average local background of $\approx 1.9$ cts. It is the only 
point-like X-ray excess in a radius $\approx$ 10\arcsec\ around 
the optical position. The background emission is mostly 
from diffuse hot gas, because the SN is relatively close 
to the star-forming galactic center. We used the Bayesian method
of \citet{KBN91} to determine whether this 
could be considered a statistically-significant detection. 
We obtain from their tables that the 90\% confidence limit 
is between 1.5 and 10.5 net counts. Hence, we conclude that 
a source coincident with the SN position is formally detected 
at the 90\% level. A longer observation is clearly needed 
to test this speculative result.

Assuming line-of-sight $N_{\rm H} = 6.2 \times 10^{20}$ 
cm$^{-2}$ \citep{K05}, and a power-law spectrum 
with $\Gamma = 2$, 5 net counts ($1.7 \times 10^{-4}$ ct s$^{-1}$)
correspond to an indicative emitted luminosity $\approx 3.5 \times 10^{37}$ 
erg s$^{-1}$ in the $0.3$--$8$ keV band. Using the 90\% confidence 
limit from Bayesian statistics, the X-ray luminosity is constrained 
between $\sim 1$--$7 \times 10^{37}$ erg s$^{-1}$. 
For an optically-thin thermal spectrum at $kT = 0.7$ keV, 
the corresponding luminosity range is 
$\sim 0.5$--$4 \times 10^{37}$ erg s$^{-1}$.

\subsection{SN\,1968D and SN\,1980K}

Discovered on 1968 February 29 \citep{Du68,W68}, 
the Type-II SN\,1968D is hosted by the SABcd galaxy NGC\,6946. 
Located at a distance of $(6 \pm 0.5)$ Mpc \citep{Kar00,Ea96},
%(e.g. Eastman et al. 1996) 
%Sharina et al. 1997; Karachentsev et al. 2000), 
NGC\,6946 has produced 9 SNe (mostly of Type II) between 1917 
and 2008---the highest number in a single galaxy.
Five of them occurred before 1970; of these, only SN\,1968D
is detected in the X-ray band (earlier SNe have also larger uncertainties 
in their optical positions). SN\,1968D   
is among the brightest X-ray sources in the galaxy (Figure 1). 
It was first recovered by \citet{Hol03} (see also Schlegel 2006), 
although it was overlooked by \citet{IK05} and was not included 
in Immler's online catalog.

We studied the X-ray colors and fluxes from each of the five  
{\it Chandra} observations, and then coadded the individual spectra 
to achieve sufficient signal-to-noise for spectral fitting. This is  
justified by the short time interval between the observations, 
compared with the age of the SN.
The coadded spectrum has 120 total counts in the $0.3$--$8$ keV band, 
and $\approx 109$ net counts.  
We rebinned the spectrum to have 9 counts per channel 
and used the Cash statistic to fit the spectrum (but we also verified 
that fits based on the $\chi^2$ statistic give almost identical results).
We used a power-law model, an optically-thin thermal plasma model, 
and both components simultaneously (Table 2), leaving 
the neutral column density free to vary. 
All three models give statistically acceptable fits. 
However, the power-law model underestimates the soft emission 
and requires an unphysically steep photon index. On the other hand, 
the thermal-plasma model slightly underestimates the emission 
above 2 keV. Thus, it is plausible that the X-ray emission is a combination 
of a non-thermal and a thermal component (Figure 2). In this case, 
we found the best-fit parameters $kT = 0.45^{+0.18}_{-0.12}$ keV 
for the gas temperature, and $\Gamma = 1.9^{+0.9}_{-1.6}$ ct s$^{-1}$ for the 
power-law photon index.
All models suggest that the absorbing column density 
is substantially larger than the Galactic line-of-sight column 
(Table 2); the emission is completely absorbed 
at energies $\la 0.7$ keV. Such large absorption leads to uncertainties 
in the estimate of the emitted luminosity. We infer  
$L_{\rm X} \sim 1$--$3 \times 10^{38}$ erg s$^{-1}$ in the $0.3$--$8$ keV 
band, and $L_{\rm X} \approx 10^{37}$ erg s$^{-1}$ in the $2$--$10$ keV 
band, from the power-law component.

We find some marginal variability but not a statistically-significant trend 
in the net count rate over the 5 observations (Table 3). 
The average observed flux from the first two datasets (from 2001--2002) 
is slightly higher than the average observed flux from 
the last three, but we cannot attribute this variability 
to a long-term decline, because of the large uncertainties.

NGC\,6946 contains various other ageing core-collapse SNe. We did not find 
any evidence of SN\,1917A and SN\,1948B. However, we have recovered SN\,1980K, 
previously detected by {\it Einstein} a month after the explosion 
\citep{Can82}, and by {\it ROSAT} 12 years later 
\citep{Sch94}. An X-ray study of SN\,1980K is also 
independently presented by \citet{Fri08}.
It has an optically-thin thermal spectrum 
%It is in the field of view of 4 of 
%the 5 {\it Chandra} observations (not in the 
%first one from 2001). It has a purely thermal spectrum 
with $kT = 0.85^{+0.04}_{-0.07}$ keV 
and an average emitted X-ray luminosity $\approx 4 \times 10^{37}$ 
erg s$^{-1}$, in the $0.3$--$8$ keV band (in fact, it has no detected 
emission above 2 keV). This is consistent with 
a value of $\approx 5 \times 10^{37}$ 
erg s$^{-1}$ in the same band, estimated from the {\it ROSAT} data 
\citep{Sch94}, after rescaling to the same adopted 
distance. We leave a more detailed discussion 
to a separate work.

\begin{deluxetable*}{lcccccc}
\tabletypesize{\scriptsize}
%\rotate
\tablecaption{Count rates from SN\,1968D from the individual observations \label{tbl-2}}
\tablewidth{0pt}
\tablehead
{
%\colhead{\,}&\colhead{\ }&\colhead{}&\colhead{}
%       &\colhead{\ }&\colhead{}&\colhead{}&\colhead{}\\[8pt]
       & \multicolumn{5}{c}{Rate ($10^{-4}$ ACIS-S ct s$^{-1}$)}\\[2pt]
       \colhead{Date}        & \colhead{SN age} &
        \colhead{$0.3$--$1$ keV}&
        \colhead{$1$--$2$ keV} &
        \colhead{$2$--$10$ keV}& 
        \colhead{$0.3$--$8$ keV} & \colhead{$0.3$--$8$ keV}  \\[2pt]
        &(days) &&&&&expected from best fit}
\startdata
2001-09-07 & 12244 & $1.0 \pm 0.5$ & $3.6\pm 0.8$  & $2.0\pm0.6$ & $6.6\pm1.1$ & $6.5$ \\[2pt]
2002-11-25 & 12688 & $1.9 \pm 0.8$ & $4.9\pm 1.3$  & $2.5\pm0.5$ & $7.7\pm1.7$ & $6.2$ \\[2pt]
2004-10-22 & 13385 & $0.6 \pm 0.5$ & $2.9\pm 1.0$  & $0.8\pm0.6$ & $4.7\pm1.3$ & $6.1$ \\[2pt]
2004-11-06 & 13400 & $1.9 \pm 0.9$ & $3.3\pm 1.1$  & $1.5\pm0.8$ & $7.1\pm1.7$ & $6.1$ \\[2pt]  
2004-12-03 & 13427 & $1.3 \pm 0.8$ & $4.8\pm 1.4$  & $0.5\pm0.5$ & $5.3\pm1.5$ & $6.2$ \\[-2pt]
\enddata
\end{deluxetable*}

\begin{figure}
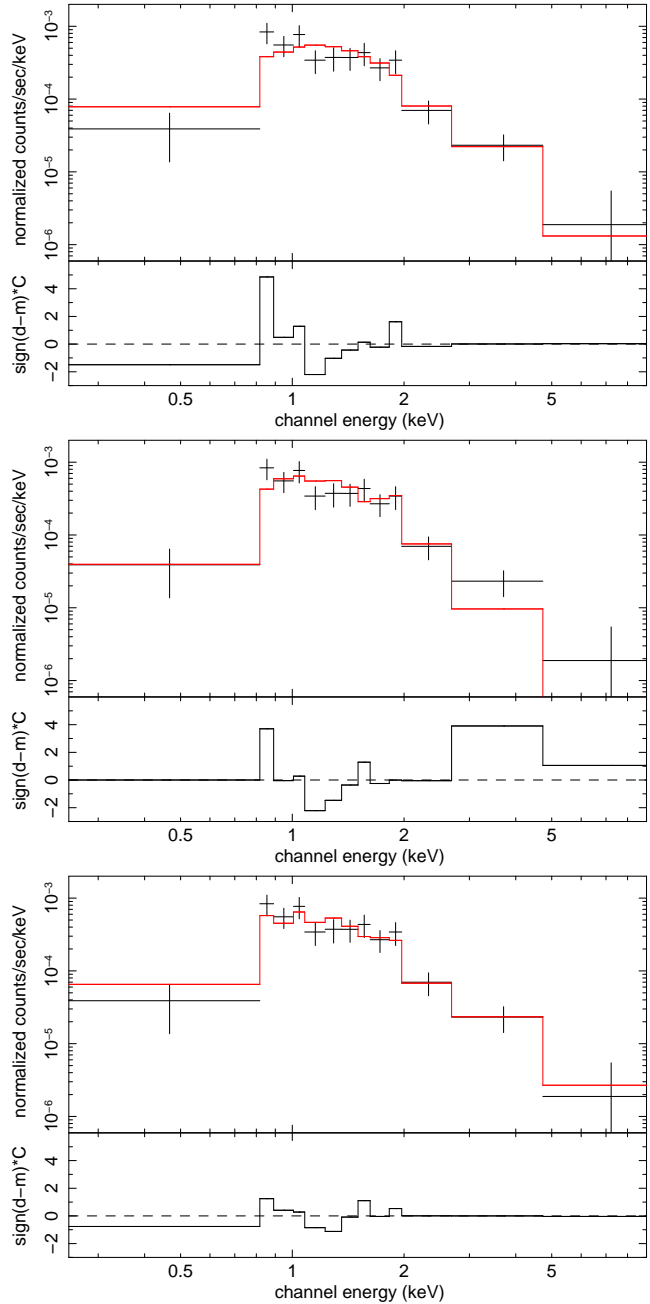

\includegraphics[angle=270,scale=.36]{fig2a.ps}\\
\includegraphics[angle=270,scale=.36]{fig2b.ps}\\
\includegraphics[angle=270,scale=.36]{fig2c.ps}
\caption{Top panel: absorbed power-law 
fit to the X-ray spectrum of SN\,1968D, with binned datapoints 
and Cash-statistic residuals.
Middle panel: same as above but for an optically-thin thermal 
plasma model.
Bottom panel: absorbed power-law plus optically thin thermal plasma 
model. The best-fit parameters and other statistics for all three 
spectral fits are in Table 2.}
\end{figure}

\section{Discussion}

Our main result is the X-ray discovery of at least one, and probably two 
historical SNe, and a detailed study of another one, 
at ages between $\approx 35$--$60$ yr. This is thought to be 
the characteristic age at which the outgoing shock reaches the interstellar 
medium, that is when the SN becomes a SN remnant \citep{IL03}. 
Before this study, SN\,1968D and SN\,1970G were the oldest confirmed 
detections \citep{Hol03,IK05}: 
we have now pushed the X-ray record to SN\,1941C. (SN 1923A has the overall 
record, having been detected in the 6-cm radio band 75 years later, 
but is not visible in X-rays; \citep{Sto06}).
Of the three new sources, only SN\,1968D had enough counts 
for spectral fitting. However, all three detections 
will be useful for deeper follow-up observations and multiband comparisons, 
and will provide constraining datapoints to model the luminosity decline in years to come. 
There are 25 historical core-collapse SNe older than SN\,1970G at distances $\la 15$ Mpc: 
16 of them are undetected in X-rays, with $L_{\rm X} \la 3 \times 10^{37}$ erg s$^{-1}$; 
3 are detected, as discussed in this paper; 
the remaining 6 have not been observed by {\it Chandra} 
or {\it XMM-Newton} yet (apart from snapshots, too short to provide meaningful 
constraints). We estimate emitted luminosities $\approx 5 \times 10^{37}$ 
erg s$^{-1}$ for SN\,1941C, $\sim$ a few $\times 10^{37}$ 
erg s$^{-1}$ for SN\,1959D, and $\approx 2 \times 10^{38}$ 
erg s$^{-1}$ for SN\,1968D, in the $0.3$--$8$ keV band.
We have also recovered another SN in NGC\,6946, 1980K, 
visible in the {\it Chandra} observations from 2002--2004, 
at a luminosity $\approx 4 \times 10^{37}$ 
erg s$^{-1}$, in the $0.3$--$8$ keV band (Soria et al., in prep.; 
see also Fridriksson et al 2008).

The optically-thin thermal-plasma emission from an ageing SN 
comes from the shocked gas between the reverse shock 
and the outgoing shell of ejecta. The X-ray luminosity
constrains the mass-loss rate of the progenitor, 
via the relation $L_{\rm X} = 4/(\pi m^2)\Lambda(T) \times (\dot{M}/v_{\rm w})^2 
\times (v_{\rm s} t)^{-1}$, where $m$ is the mean mass per particle, 
%($2.1 \times 10^{24}$ g for a H$+$He plasma), 
$\Lambda(T)$ is the cooling 
function at temperature $T$, $\dot{M}$ the mass-loss rate of the progenitor, 
$v_{\rm w}$ the stellar wind speed and $v_{\rm s}$ the speed of the outgoing shock 
\citep{Im02,IL03}. Hence, the X-ray luminosity 
of a SN at an age $t$ depends on the wind properties of the progenitor 
at an age $t v_{\rm s}/v_{\rm w} \sim 10^3 t$ before the explosion. 
For our fitted luminosity of SN\,1941C, using the 
numerical values from \citet{IK05}, we infer a mass-loss rate 
$\approx 5 \times 10^{-5} M_{\odot}$ yr$^{-1}$ at $\approx 55,000$ yr 
before the SN. For SN\,1968D (Table 2), the inferred mass-loss rate is  
$\approx 8 \times 10^{-5} M_{\odot}$ yr$^{-1}$ at $\approx 30,000$ yr 
before the SN; for SN\,1980K, $\dot{M} \approx 3 \times 10^{-5} M_{\odot}$ 
yr$^{-1}$ at a time $\approx 20,000$ yr. More likely, these 
should be considered as an order of magnitude estimates.
The X-ray luminosity and mass-loss rate of SN\,1968D 
suggest that this SN is evolving to a remnant similar to Cas A 
\citep{Dun03,Hw04}. The similarity between 
SN\,1968D and Cas A was also noted from radio observations 
\citep{Hym95}. 

%The radio behaviour of SN\,1968D was also 
%found to be similar to that of SN\,1980K 
%\citep{Hym95}.
%, which we have also recovered in the {\it Chandra} 
%observations from 2002--2004, at a luminosity $\approx 4 \times 10^{37}$ 
%erg s$^{-1}$, in the $0.3$--$8$ keV band (Soria et al., in prep.). 
In Type-II SNe, the synchrotron radio emission is produced by 
electrons accelerated at the outgoing shock front, 
where the ionized circumstellar wind is compressed, 
and the magnetic field locked in it is amplified
\citep{Che82a,Che82b,Wei86,Wei92}.
Thus, the radio luminosity is also a function of $\dot{M}/v_{\rm w}$ 
\citep{Wei86}. 
However, a single radio measurement at late times is generally 
not sufficient to determine the mass-loss rate; 
one has to know other parameters 
such as the density profile of the ejecta and 
the optical depth and flux at early times (e.g., 1 day 
after the explosion). Of the four X-ray SNe discussed 
here, 1980K is the only one for which a detailed 
radio lightcurve is available \citep{Mon98}. From a detailed 
modelling of their data, \citet{Mon98} infer a mass-loss rate 
$\dot{M} \approx 2 \times 10^{-5} M_{\odot}$ yr$^{-1}$
for the first 10 yrs after the SN, possibly declining 
to $\approx 10^{-5} M_{\odot}$ yr$^{-1}$ after that. 
This is in good agreement with the X-ray estimate.

We found no published radio fluxes or limits for 
SN\,1941C. There are three upper limits 
for SN\,1959D, at $\approx 26$--$27$ yrs after the explosion 
\citep{Eck02}, and a radio detection for SN\,1968D, 
$\approx 26$ yrs after the event \citep{Hym95}.
In the absence of early-time data or detailed 
radio lightcurves, it may still be possible to 
estimate $\dot{M}$, by using characteristic parameters 
from another SN assumed as a standard template; for example,
\citet{Eck02} suggest an empirical formula based 
on the behaviour of SN\,1979C. Using their formula, 
the stellar mass-loss rate for SN\,1959D would be 
$< 6 \times 10^{-6} M_{\odot}$ yr$^{-1}$, 
and $\approx 7 \times 10^{-6} M_{\odot}$ yr$^{-1}$ 
for SN\,1968D, an order of magnitude less than estimated from 
the X-ray luminosity. However, we suspect that 
Eck's scaling may under-estimate 
some mass-loss rates. For instance, 
the same relation gives $\dot{M} \approx 6 \times 10^{-6} M_{\odot}$ 
yr$^{-1}$ for SN\,1980K, three times less than 
calculated from detailed radio modelling; 
for SN\,1970G, Eck's formula implies 
$\dot{M} \approx 5 \times 10^{-6} M_{\odot}$ yr$^{-1}$ 
while the X-ray estimate is  $\approx (2$--$3) \times 
10^{-5} M_{\odot}$ yr$^{-1}$ \citep{IK05}.
A more accurate estimate of the mass-loss rate 
is important not only to determine the type 
of progenitor star, but also to constrain the time 
of transition between the SN and SN remnant phase: 
the outgoing shock overtakes the stellar wind bubble 
and reaches the interstellar medium after a characteristic 
time $\sim 10^{-3} (\dot{M})^{-1}$ yr \citep{Wei86}.
We leave a more detailed comparison of radio- and X-ray-inferred 
rates in SN\,1980K and other X-ray SNe to further work.

The presence of a power-law-like emission component above 2 keV in
SN\,1968D suggests that there may be a contribution from a young
pulsar and its wind nebula. We estimate an average power-law luminosity $\approx 10^{37}$
erg s$^{-1}$, in the $2$--$10$ keV band. Interestingly, this is
comparable to the luminosity of Crab-like systems \citep{P02}, 
and in particular the Crab itself and PSR\,B0540$-$69 \citep{Ser04}.  
Therefore, deep X-ray observations of SNe older than
about 30 years are optimally suited for the discovery of very young
pulsars, and hence to constrain the statistics of pulsar formation,
as well as the distribution of their spins at birth.

%Pulsar emission may be absorbed for a few decades 
%after the explosion, and may become visible at later times. 
%It may become the dominant source of X-rays, through 
%a wind nebula and direct magnetospheric emission (eg Crab) 
%So, studying SNe older than 30 yrs is crucial to constrain the statistics 
%of pulsar formation, and the initial distribution of the spin parameter 
%which determines the energy loss rate and therefore 
%the X-ray luminosity.
%The emission component from a pulsar is a power-law 
%with $\Gamma \approx 2$, which should be easy to distiguish 
%from the softer thermal-plasma component emitted by the shocked gas.

\acknowledgments

We are grateful to S. Immler and D. Pooley for discussions 
and collaborations in this project, and to the anonymous referee 
for bringing up the issue of radio estimates of mass-loss rates. 
We thank E. M. Schlegel and S. S. Holt for pointing out their 
X-ray discovery of SN\,1968D (missing from other catalogs),  
and J. K. Fridriksson and A. K. H. Kong for drawing our attention 
to their new X-ray study of NGC\,6946 including SN\,1980K.
RS acknowledges support from a Leverhulme Fellowship, 
and also from Tsinghua University (China) during part 
of this project.

%\clearpage

%\clearpage

%\clearpage

%\clearpage
%\pagebreak

%\clearpage

\end{document}